\def\rfr#1{eq. (\ref{#1})}

\def\Rfr#1{Eq. (\ref{#1})}

\def\virg#1{``#1''}

 \def\bb{\bibitem}
\def\eqi{\begin{equation}}
\def\eqf{\end{equation}}
\def\eqia{\begin{eqnarray}}
\def\eqfa{\end{eqnarray}}
\def\rp#1#2{\frac{#1}{#2}}
 \def\lb#1{\label{#1}}

\def\bds#1{\boldsymbol{#1}}

\documentclass[usenatbib]{mn2e}
\usepackage{amsmath}
\usepackage{amscd}
\usepackage{amssymb}
\usepackage{latexsym}
\usepackage{graphicx}
\usepackage{epsfig}

\title[Galactic orbital motions  in CDM, MOND and MOG]{Galactic  orbital motions  in the Dark Matter, MOdified Newtonian Dynamics and MOdified Gravity scenarios}

\author[L. Iorio]{
L. Iorio$^{1}$\thanks{E-mail:
lorenzo.iorio@libero.it}\\
$^{1}$INFN-Sezione di Pisa, Viale Unit$\grave{\rm a}$ di Italia 68, 70125, Bari (BA), Italy
}

\begin{document}

\date{Accepted 2009 September 28. Received 2009 September 23; in original form 2009 March 12}


\maketitle

\label{firstpage}

%
%
%
%
%
%
%
%
%
%
%
%
\begin{abstract}
  We simultaneously integrate in a numerical way the equations of motion of both the Magellanic Clouds (MCs) in the MOdified Newtonian Dynamics (MOND), MOdified Gravity (MOG) and Cold Dark Matter (CDM) frameworks for $-1\leq t\leq 1$ Gyr in order to see if, at least in principle, it is possible to discriminate between them. Since the Large Magellanic Cloud (LMC) and the Small Magellanic Cloud (SMC) are at distances  of approximately 50-60 kpc from the center of the Milky Way (MW), they are ideal candidates to investigate the deep MOND regime
occurring when the characteristic MOND acceleration $A_0=1.2\times 10^{-10}$ m s$^{-2}$ is larger than the internal acceleration $A$ of the system considered: indeed, the Newtonian baryonic accelerations $A_{\rm N}$ involved are about $0.02-0.03 A_0$ for them.
It turns out that CDM, MOND and MOG yield, in fact, different trajectories.
In MOND also the External Field Effect (EFE) $A_{\rm ext}$ must, in principle, be considered. Since for MW $A_{\rm ext}\approx 0.01 A_0$, with a lingering uncertainty, we consider both the cases $A_{\rm ext}\ll A_{\rm N}, A_{\rm ext}\ll A_0$   and $A_{\rm ext}=A_{\rm N}, A_{\rm ext}\ll A_0$. We also investigate the impact of the current uncertainties in the velocity components of  MCs on their motions in the theories considered. In modeling the mutual interaction between the clouds and the dynamical friction (in CDM and MOND) we use for the masses of MCs the total (baryonic $+$ dark matter) values, dynamically inferred,
 in CDM, and the smaller ones (baryonic), coming from direct detection of visible stars and neutral gas, in MOND and MOG.

\end{abstract}



\section{Introduction}\lb{Intro}
In many astrophysical systems like, e.g., spiral galaxies and clusters of galaxies a discrepancy between the observed kinematics of some of their components and the predicted one on the basis of the Newtonian dynamics and the matter directly detected from the emitted electromagnetic radiation (visible stars and gas clouds) was present  since the pioneering studies by\footnote{He postulated the existence of undetected, baryonic matter; today, it is believed that the hidden mass is constituted by non-baryonic, weakly interacting particles in order to cope with certain issues pertaining galaxy formation and primordial nucleosynthesis \citep{Gond}.}  \citet{Zwi} on the Coma cluster, and by \citet{Bos} and \citet{Rub} on spiral galaxies. More precisely, such an effect shows up in the galactic  velocity rotation curves \citep{flatt,flattona} whose typical pattern after a few kpc from the center differs from the Keplerian $1/\sqrt{r}$ fall-off expected from the usual Newtonian dynamics applied to the electromagnetically-observed matter.

As a possible solution of this puzzle, the existence of non-baryonic, weakly-interacting Cold Dark (in the sense that its existence is indirectly inferred only from its gravitational action, not from emitted electromagnetic radiation) Matter (CDM) was proposed to reconcile the predictions with the observations \citep{Rub83} in the framework of the standard gravitational physics; for a general review on the CDM issue see, e.g, \citet{Kha}, while for the distribution of CDM in galaxies, see, e.g., \citet{Sal}. To be more definite, let us focus on the Milky Way (MW) and adopt a very widely  used model of its gravitational potential $U$. It consists of the standard Miyamaoto-Nagai  disk
\citep{Miy75}
\eqi U_{\rm disk} = -\rp{\xi GM_{\rm disk}}{\sqrt{x^2+y^2+\left(k+\sqrt{z^2+b^2}\right)^2}}, \lb{MiyaNaga}\eqf
the \citet{Plum11} bulge
\eqi U_{\rm bulge} = -\rp{GM_{\rm bulge}}{r+c}, \lb{Plumm}\eqf
and the logarithmic CDM halo  by \citet{BinTrem87}
\eqi U_{\rm halo} = v^2_{\rm halo}\ln\left(r^2+d^2\right),\ ({\rm spherical\ halo})\lb{BinTrem}\eqf
with  \citep{Law,Will09}
$\xi=1$, $k=6.5$ kpc, $b=0.26$ kpc, $c=0.7$ kpc, $v_{\rm halo}=114$ km s$^{-1}$, $d=12$ kpc. The masses of the disk and the bulge used by \citet{Law} are those by
\citet{John99}, i.e. $M_{\rm disk}=1\times 10^{11}$ M$_{\odot}$, $M_{\rm bulge}=3.4\times 10^{10}$ M$_{\odot}$  yielding a total baryonic mass of $M=1.34\times 10^{11}$ M$_{\odot}$; however, such a value is almost twice the most recent estimate ($M=6.5\times 10^{10}$ M$_{\odot}$) by \citet{McG} who includes the gas mass as well and yield $M_{\rm disk}=2.89\times 10^{10}$ M$_{\odot}$ and $M_{\rm bulge}=2.07\times 10^{10}$ M$_{\odot}$.   \citet{Xue} yield a total baryonic mass of $M=6.5\times 10^{10}$ M$_{\odot}$ as well; they use a different bulge+disk+CDM halo model of the Galaxy with $M_{\rm disk}=5\times 10^{10}$ M$_{\odot}$
and $M_{\rm bulge}=1.5\times 10^{10}$ M$_{\odot}$, as in \citet{Smith07}.
The model of \rfr{MiyaNaga}-\rfr{BinTrem}, with the parameters' values by \citet{Law} and \citet{John99}, has been recently used by \citet{Will09}
to study the motion of the \citet{Gril} tidal stellar stream at Galactocentric distance of $r\lesssim 16-18$ kpc; \citet{Read} used it to study the motion of the tidal debris of the Sagittarius dwarf at 17.4 kpc from the center of  MW.  More specifically, the CDM halo model of \rfr{BinTrem} corresponds to a CDM halo mass
\eqi M_{\rm halo}=\rp{2 v_{\rm halo}^2 r^3}{G(r^2+d^2)},\lb{massina}\eqf
so that
\eqi M_{\rm halo}(r=60\ {\rm kpc})= 3.5\times 10^{11}\ {\rm M}_{\odot},\eqf
in agreement with the value by \citet{Xue}
\eqi  M_{\rm halo}(r=60\ {\rm kpc})= (4.0\pm 0.7)\times 10^{11}\ {\rm M}_{\odot}.\lb{massaccia}\eqf
Concerning $v_{\rm halo}$,  other authors report different values for it; e.g., \citet{Read} use $v_0=175$ km s$^{-1}$, where $v_0^2=2v_{\rm halo}^2$, so that $v_{\rm halo}=124$ km s$^{-1}$ for them, while \citet{John99} yield the range $140-200$ km s$^{-1}$ for their $v_{\rm circ}=\sqrt{2}v_{\rm halo}$ which maps into $70$ km s$^{-1}$ $\leq v_{\rm halo}\leq 141$ km s$^{-1}$. However, it must be noted that values of $v_{\rm halo}$ too different from 114 km s$^{-1}$ would destroy the agreement of \rfr{massina} with the value of \rfr{massaccia}.

Oppositely, it was postulated that the Newtonian laws of gravitation have to be modified on certain acceleration scales to correctly account for the observed anomalous kinematics of such astrophysical systems without resorting to still undetected exotic forms of matter.
One of the most phenomenologically successful modifications of the inverse-square Newtonian acceleration $A_{\rm N}$, mainly with respect to spiral galaxies, is the MOdified Newtonian Dynamics (MOND) \citep{Mil83a,Mil83b,Mil83c} which  postulates that
for systems  experiencing total gravitational acceleration  $A \ll A_0$, with \citep{Bege} \eqi A_0= (1.2\pm 0.27)\times 10^{-10} \ {\rm m\ s}^{-2},\eqf
\eqi \bds A\rightarrow \bds A_{\rm MOND}=-\rp{\sqrt{A_0GM}}{r}\bds{\hat{r}}.\lb{MOND}\eqf   More precisely, it holds \eqi A = \rp{A_{\rm N}}{\mu(X)},\ X\equiv\rp{A}{A_0};\lb{appromond}\eqf
$\mu(X)\rightarrow 1$ for $X\gg 1$, i.e. for large accelerations (with respect to $A_0$), while $\mu(X)\rightarrow X$ yielding \rfr{MOND} for $X\ll 1$, i.e. for small accelerations (again, with respect to $A_0$).
The most widely used forms for the interpolating function $\mu(X)$ are  the \virg{standard}
 \citep{joint}
\eqi \mu(X) = \rp{X}{\sqrt{1+X^2}},\lb{mu2}\eqf
and the simpler
\citep{Fam}
\eqi \mu(X) = \rp{X}{1+X}.\lb{mu1}\eqf
It recently turned out that \rfr{mu1} yields better results in fitting the terminal velocity curve of  MW, the rotation curve of the standard external galaxy NGC 3198 \citep{Fam,kazzo,scassa} and of a sample of 17 high
surface brightness, early-type disc galaxies \citep{Noor};
\rfr{appromond} becomes
\eqi A = \rp{A_{\rm N}}{2}\left(1+\sqrt{1+\rp{4 A_0}{A_{\rm N}}}\right)\lb{oleee}\eqf
with \rfr{mu1}. \Rfr{appromond} strictly holds for co-planar, spherically and axially symmetric mass distributions \citep{Brada}; otherwise, the full  modified (non-relativistic) Poisson equation   \citep{joint}
\eqi \bds\nabla\bds\cdot\left[\mu\left(\rp{|\bds\nabla U|}{A_0}\right)\bds\nabla U\right]=4\pi G\rho\eqf
must be used. Attempts to yield a physical foundation to MOND, especially in terms of a relativistic covariant theory, can be be found in, e.g., \citet{joint,Bek,Far,Zha07}; for recent reviews of various aspects of the MOND paradigm, see \citet{San02,Bek06,Mil08}.
The compatibility of MOND with Solar System data has been investigated by \citet{Mil83a,Tal,Ser,Magu,San06,Ior,Ior09,Mil09}.
Generally speaking, many theoretical frameworks have been set up to yield a $1/r$ acceleration term able to explain the observed dynamics of astrophysical systems; for example, those encompassing a logarithmic extra-potential \citep{Cad,Fab}
\eqi U=C\ln\left(\rp{r}{r_s}\right),\eqf
where $C$ and $r_s$, a length scale, are fit-for parameters.
For other modified models of gravity used to explain, among other things, the galactic rotation curves without resorting to CDM see, e.g., \citet{Capo,Fri,Mof}.

The MOdified Gravity (MOG) \citep{Mof} is a fully covariant
theory of gravity which is based on the existence of a massive vector field
coupled universally to matter. The theory yields a Yukawa-like modification of
gravity with three constants which, in the most general case, are running; they
are present in the theory's action as scalar fields which represent the
gravitational constant, the vector field coupling constant, and the vector
field mass.   An approximate solution of the
MOG field equations \citep{cazMOF} allows to compute their values as functions of the
source's mass. The resulting Yukawa-type modification of the
inverse-square Newton's law in the gravitational field of a central mass $M$ is
\eqi \bds A_{\rm MOG} = -\rp{G_{\rm N}M}{r^2}\left\{1+\alpha\left[1-\left(1+\mu r\right)\exp\left(-\mu r\right)\right]\right\}\bds{\hat{r}},\lb{Mog}\eqf
with\footnote{\citet{Mof} used the equivalent notation $E\rightarrow C_1^{'}$ and $D\rightarrow C_2^{'}$.}
\begin{eqnarray}
  \alpha &\simeq & \rp{M}{\left(\sqrt{M}+E\right)^2}\left(\rp{G_{\infty}}{G_{\rm N}}-1\right), \\
  \mu &\simeq & \rp{D}{\sqrt{M}},
\end{eqnarray}
where  $G_{\rm N}$ is the Newtonian gravitational constant  and
\begin{eqnarray}
  G_{\infty} &\simeq & 20 G_{\rm N}, \\
  E &\simeq & 25,000\sqrt{{\rm M}_{\odot}}, \\
  D &\simeq & 6,250\sqrt{{\rm M}_{\odot}}\ {\rm kpc}^{-1}.
\end{eqnarray}
Such values have been obtained by \citet{Mof,cazMOF} as a result of the fit of the velocity rotation curves of some
galaxies in the framework of the searches for an explanation of the rotation curves of galaxies without resorting to CDM.
The validity of \rfr{Mog} in the Solar System has been recently questioned in \citet{IorMOG}.
For \citep{McG} $M=6.5\times 10^{10}$ M$_{\odot}$, we have
\begin{eqnarray}
  \alpha &\simeq & 16 \\
  \lambda &=& \rp{1}{\mu} \simeq \ 41\ {\rm kpc}.
\end{eqnarray}

Traditionally, the phenomenology of both MOND and CDM paradigms is based on the electromagnetically detected matter (stars and gas clouds) at no more than about 20 kpc; in view of the use by \citet{Clew04} and \citet{Xue}  of several recently discovered Blue Horizontal-Branch (BHB) stars as kinematical tracers at large radii ($r\approx 60-130$ kpc), it makes now sense to look at the remote periphery of the Galaxy as well to try to test CDM and alternative models of gravity.
In this paper we wish to investigate the orbits of test particles at Galactocentric distances $r>20$ kpc, i.e. in the deep MONDian regime; we will use the Magellanic Clouds (MCs) moving at 50-60 kpc from the center of  MW.  We will extend our analysis also to MOG  and to the action of CDM itself as well  to see if our approach is able, at least in principle, to discriminate between them;  for another attempts on galactic scales, based on the escape speed in the solar neighborhood, see also \citet{Zha}. At so large Galactocentric distances many complications arising from an accurate modeling of the realistic distribution of mass can be avoided, both in MOND/MOG and in CDM frameworks.
Moreover, \citet{Gar96}, \citet{Yos03} and \citet{Con06}
demonstrated that the position of the Magellanic Stream (MS)
follows the orbits of  MCs. Therefore, it is interesting to
compare the path of MS with the orbits
predicted by CDM, MOND and MOG.
Thus, it is hoped that our results will encourage more quantitative and detailed studies on MOND and MOG applied to such systems; for numerical investigations on the problem of the formation of cosmological structures and galactic evolution, see \citet{Knebe,Iran1,Cosmo,Tir,Tir2,Iran2}.

\section{Motions in CDM, MOND and MOG: the Magellanic Clouds}\lb{grandeee}
 Concerning MOND and MOG, we will consider a central body with the same mass \citep{McG} $M\approx 6.5\times 10^{10}$M$_{\odot}$ of the total baryonic component of  MW and a test particle distant several tens kpc from it, acted upon by the putative MOND/MOG gravitational fields of $M$. Such large distances
allow to neglect the details of the real mass distribution which may become relevant in MOND at closer distances \citep{Read,Nipo}. To preliminarily test our approximation we applied \rfr{oleee} to the Sagittarius dwarf galaxy ($r=17.4$ kpc) and confronted the numerically integrated orbital sections in the coordinate planes of the trajectory to those obtained by \citet{Read} by using a non-pointlike baryonic potential (upper panel of Figure 2 in \citep{Read}); we used the same integration interval of $-1$ Gyr $\leq t\leq 1$ Gyr and the same baryonic mass ($M = 1.2\times 10^{11}$M$_{\odot}$) by \citet{Read}. It turns out that we were successful in reproducing the orbital sections by \citet{Read}; thus, we are confident of the validity of our approximation for the larger Galactocentric distances we will use in the following analysis.

Another issue which, in principle, should be taken into account in MOND is the so-called External-Field-Effect (EFE); it may become relevant with cluster of galaxies \citep{Clu}.
According to, e.g.,  \citet{San02,Zha,Angus08},
\eqi \mu\left(\rp{|\bds A_{\rm ext}+\bds A|}{A_0}\right)A=A_{\rm N},\lb{effe}\eqf
where $A_{\rm N}$ is the Newtonian acceleration of the  system alone, $A$ is its  total internal acceleration, while $A_{\rm ext}$ denotes the acceleration induced by any external field.
By using the simpler form of \rfr{mu1} for $\mu$, one approximately obtains  from \rfr{effe}
\eqi A\approx \rp{A_{\rm N}}{2}\left[1-\rp{A_{\rm ext}}{A_{\rm N}} + \sqrt{\left(1-\rp{A_{\rm ext}}{A_{\rm N}}\right)^2
+\rp{4A_0}{A_{\rm N}}\left(1+\rp{A_{\rm ext}}{A_0}\right)}
\right].\lb{mega}\eqf
For $A_0\rightarrow 0$, $A\rightarrow A_{\rm N}$, as expected.
For $A_{\rm ext}\rightarrow 0$, i.e. $A_{\rm ext}\ll A_0$ and $A_{\rm ext}\ll A_{\rm N}$, one has  $A \rightarrow$  \rfr{oleee}.
For
\eqi\rp{A_{\rm ext}}{A_0}\ll 1\eqf
only,
the total acceleration becomes
\eqi A\approx\rp{A_{\rm N}}{2}\left[1-\rp{A_{\rm ext}}{A_{\rm N}}+\sqrt{\left(1-\rp{A_{\rm ext}}{A_{\rm N}}\right)^2+\rp{4A_0}{A_{\rm N}}}\right],\lb{fuffo}\eqf
while for \eqi \rp{A_{\rm ext}}{A_{\rm N}}\approx 1\eqf
only,
it is
\eqi A\approx \sqrt{A_{\rm N}A_0\left(1+\rp{A_{\rm ext}}{A_0}\right)}.\eqf
Interestingly, if
\eqi\rp{A_{\rm ext}}{A_{\rm N}}\approx 1,\ \rp{A_{\rm ext}}{A_0}\ll 1,\eqf
then
\eqi A\approx \sqrt{A_{\rm N}A_0}=\rp{\sqrt{GMA_0}}{r}.\lb{stracaz}\eqf
In the case of  MW, it is very difficult to reliably assess the external field because it may be due to several factors like, e.g., the Large Scale Structure and the Great Attractor region ($A_{\rm ext}/A_0=0.01$), but also the galaxy M31 Andromeda, at 800 kpc from  MW, and the Coma and Virgo clusters, whose field are time-varying, may play a role. For a discussion see \citet{WuMag}.  In view of the lingering uncertainty of $A_{\rm ext}$, in the following we will use \rfr{oleee}; however, we will also investigate the case in which $A_{\rm ext}=A_{\rm N},\ A_{\rm ext}\ll A_0$ because it may occur in  MW at the large Galactocentric distances considered here.

As a concrete example of motion in deep MOND regime ($A_{\rm N}/A_0\approx 0.03-0.02$), let us consider both MCs; the Large Magellanic Cloud (LMC) is at 49.4 kpc from the center of  MW (Galactic Center, GC), while the Small Magellanic Cloud (SMC) is located at 59 kpc from GC. LMC and SMC's
Galactocentric cartesian coordinates and velocities  \citep{Kalli,WuMag} are in Table  \ref{LMCcoor} and Table  \ref{SMCcoor}.
 \begin{table}
\caption{Large Magellanic Cloud (LMC):  coordinates \citep{Kalli,WuMag}, in kpc, and velocity components \citep{WuMag}, in km s$^{-1}$, of LMC  in a Galactocentric rest frame $\{X,Y,Z\}$ with the $Z-$axis pointing toward the Galactic north pole, the $X-$axis
pointing in the direction from the Sun to the Galactic center, and
the $Y-$axis pointing in the direction of the Sun's Galactic rotation \citep{Kalli,Besla}. They yield $r=49.5$ kpc, $v=378$ km s$^{-1}$.
The uncertainties in the coordinates can be neglected \citep{Cioni}.}
\centering
\begin{tabular}{@{}lll@{}}
\hline
$X_0 = -0.8$  & $Y_0 = -41.5$ & $Z_0 = -26.9$ \\
%
%
$\dot X_0 = -86\pm 12$ & $\dot Y_0 =-268\pm 11$  & $\dot Z_0 =252\pm 16$\\
\hline
\end{tabular}\label{LMCcoor}
\end{table} %
\begin{table}
\caption{Small Magellanic Cloud (SMC):  coordinates \citep{Kalli,WuMag}, in kpc, and velocity components \citep{WuMag}, in km s$^{-1}$, of SMC in a Galactocentric rest frame $\{X,Y,Z\}$ with the $Z-$axis pointing toward the Galactic north pole, the $X-$axis
pointing in the direction from the Sun to the Galactic center, and
the $Y-$axis pointing in the direction of the Sun's Galactic rotation \citep{Kalli,Besla}.  They yield $r=58.9$ kpc, $v=301$ km s$^{-1}$.
The uncertainties in the coordinates are negligible \citep{Cioni}.}
\centering
\begin{tabular}{@{}lll@{}}
\hline %
$X_0 = 15.3$ &$Y_0 = -36.9$  &$Z_0 = -43.3$\\
%
%
$\dot X_0 = -87\pm 48$ & $\dot Y_0 =-247\pm 42$  & $\dot Z_0 =149\pm 37$\\
\hline
\end{tabular}\label{SMCcoor}
\end{table} %
It can be noted that the velocity components of LMC are uncertain at more than $4-14\%$. The situation for the position components is much better since they are known with uncertainties in the range $0.1-1\%$, as it results from the analysis of the tip of the red giant
branch (TRGB) applied to MCs by  \citet{Cioni}; thus, we will neglect them in the following. Also for SMC the uncertainty in the position components is negligible \citep{Cioni}, while the velocity components are known at $10\%$.

We simultaneously integrated in a numerical way the equations of motion of both MCs in MOND, MOG and CDM  by using the initial conditions of Table \ref{LMCcoor} and of Table   \ref{SMCcoor} for $-1\leq t\leq 1$ Gyr.
In addition to the main pull due to MW, we also included the mutual attractions of MCs   and the effect of the dynamical friction  due to their motion through the Galactic dark halo \citep{BinTrem87}; the mutual dynamical friction  was neglected \citep{Kalli}.
Concerning the pull by LMC on SMC, we modeled its action in Newtonian dynamics from a \citet{Plum11}-type potential  \citep{Kalli}
\eqi U_{\rm LMC}=\rp{Gm_{\rm LMC}}{\sqrt{(x-x_{\rm LMC})^2+(y-y_{\rm LMC})^2+(z-z_{\rm LMC})^2+ K^2_{\rm LMC}}},\eqf
with $K_{\rm LMC}=3$ kpc. In MOND, since the acceleration imparted by LMC on SMC is of the order of about 0.05$A_0$, we adopted \rfr{MOND} with $M\rightarrow m_{\rm LMC}$ and $r=\sqrt{(x-x_{\rm LMC})^2+(y-y_{\rm LMC})^2+(z-z_{\rm LMC})^2+K^2_{\rm LMC}}$, while in MOG we used \rfr{Mog} with $M\rightarrow m_{\rm LMC}$ and $r=\sqrt{(x-x_{\rm LMC})^2+(y-y_{\rm LMC})^2+(z-z_{\rm LMC})^2+K^2_{\rm LMC}}$. An analogous expression for the pull by SMC on LMC holds; in this case, $K_{\rm SMC} =2$ kpc \citep{Kalli}.
The dynamical friction experienced by, say, SMC in going through the dark halo of the Galaxy has been modelled, in CDM, as
\eqi\bds D=-\rp{\bds v}{t_{\rm fric}},\eqf
with \citep{Kalli,drag} \eqi t^{-1}_{\rm fric}\approx 0.428\ln\Lambda \rp{Gm_{\rm SMC}}{r^2 v},\lb{dynappr}\eqf where the Coulomb logarithm $\ln\Lambda\approx 3$ \citep{BinTrem87}.
We also included the mutual dynamical friction experienced by SMC when its distance from LMC gets smaller than 15 kpc \citep{Bekki,Kalli} by replacing in \rfr{dynappr} $\ln\Lambda=3$ with $\ln\Lambda_{\rm LS}=0.2$ and $r$ with $r_{\rm mutual}$. The dynamical friction plays a non-negligible role also in several astrophysical systems in the framework of MOND  \citep{Ciot,Salcedo,Nipoti}; in our case, we model it  by assuming \citep{Ciot,Nipoti}\eqi \rp{t^{\rm MOND}_{\rm fric}}{t^{\rm N}_{\rm fric}}=\rp{\sqrt{2}}{1+\rp{A}{A_{\rm N}}}\approx \rp{\sqrt{2}}{1+\sqrt{\rp{A_0}{GM}}r}.\eqf  Since a model of the dynamical friction has not yet been developed in the framework of MOG, we did not include it.

Concerning  MCs' masses entering both their mutual interactions and the dynamical friction, for consistency reasons we adopted the total (baryonic $+$ dark matter) values $m_{\rm LMC}=2\times 10^{10}$M$_{\odot}$  \citep{Kalli,drag,Sch,Gar96}, coming from radial velocities
of several of the oldest star clusters in  LMC lying well
beyond 6 kpc from its center, and\footnote{This typical value has been chosen by \citet{Kalli} after an examination of the values coming from observations of carbon stars \citep{Har} and planetary nebul{\ae} \citep{Dop}, and from a virial analysis of the kinematics of thousands of red giant stars in
SMC \citep{Zar}.} $m_{\rm SMC}=3\times 10^{9}$M$_{\odot}$  \citep{Kalli,drag} when integrating the CDM model. Instead, we used  the smaller, baryonic  values $m_{\rm LMC}=(2.7+0.5=3.2)\times 10^{9}$M$_{\odot}$  (visible disk $+$  neutral gas \citep{Kim}) \citep{van,van08} and $m_{\rm SMC}=(3.1+5.6 = 8.7)\times 10^{8}$M$_{\odot}$ (total stellar mass $+$  neutral gas) \citep{van08} for MOND and MOG. For a recent discussion of the methods employed to obtain such figures and of other results, see \citep{van08}.
\subsection{The Large Magellanic Cloud}
\begin{figure}
\centerline{
\vbox{
\epsfysize= 5.75 cm\epsfbox{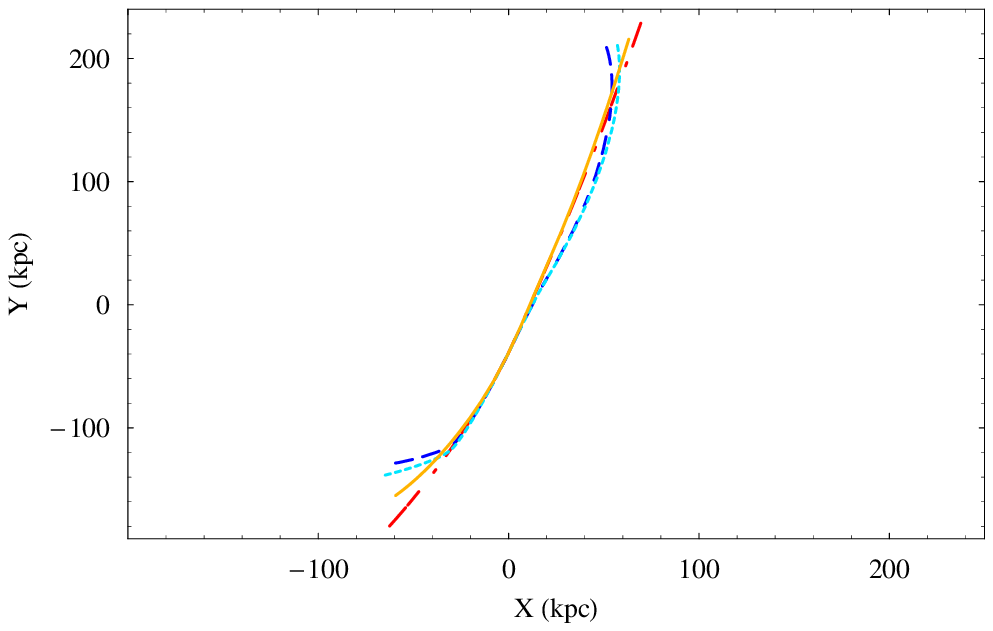}
\epsfysize= 5.75 cm\epsfbox{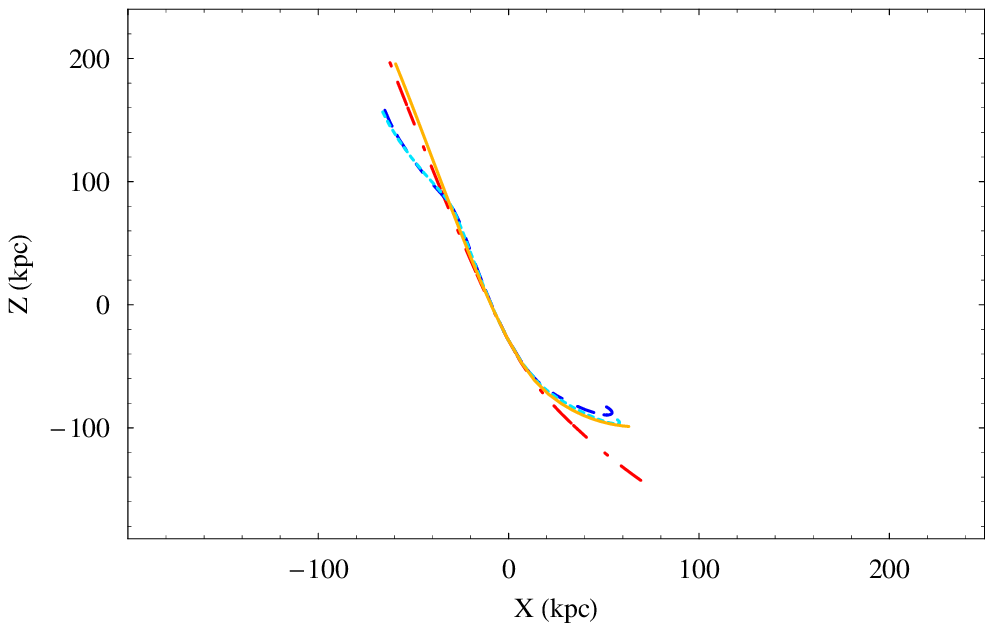}
\epsfysize= 5.75 cm\epsfbox{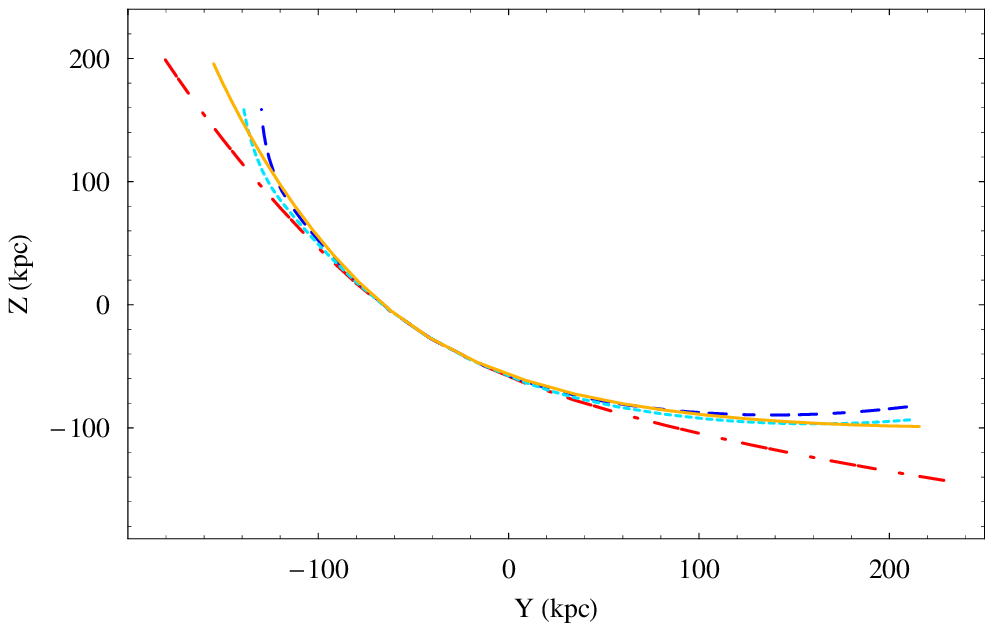}
      }
%
%
%
}
\caption{ {Sections in the coordinate planes of the numerically integrated trajectories of  LMC experiencing: a) The Newtonian acceleration with CDM (red dash-dotted  line) b)  The MOND acceleration with $\mu=X/(1+X)$   (blue dashed  line) c) The MOND acceleration with $\mu=X$   (light blue dotted line) d) The MOG acceleration (yellow continuous  line). The central values of the initial conditions of Table  \ref{LMCcoor} have been used. For the baryonic masses of  MW's bulge and  disk  we used the values by \protect{\citet{McG}}, with a total baryonic mass of $M=6.5\times 10^{10}$ M$_{\odot}$.  For the masses of MCs, entering their mutual interactions and the dynamical friction, both modelled in the present integration, the total  values (baryonic +  dark matter) have been adopted for CDM, while those encompassing only baryonic components have been used for MOG and MOND. The time span of the integration is $-1\leq t\leq 1$ Gyr.} \label{LMC_planes}
}
\end{figure}

\begin{figure}
\centerline{
\vbox{
\epsfysize= 4.75 cm\epsfbox{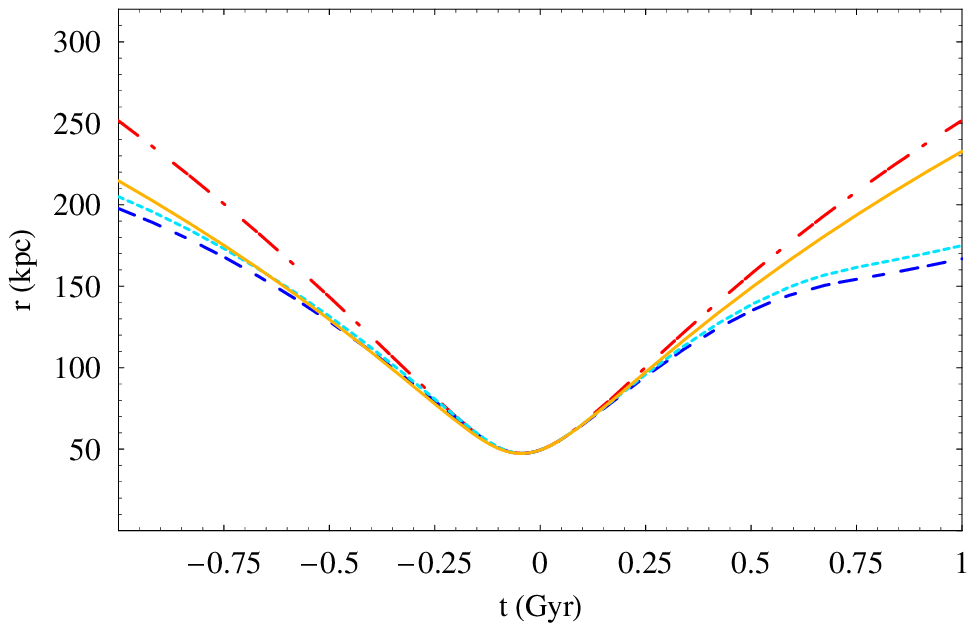}
\epsfysize= 4.75 cm\epsfbox{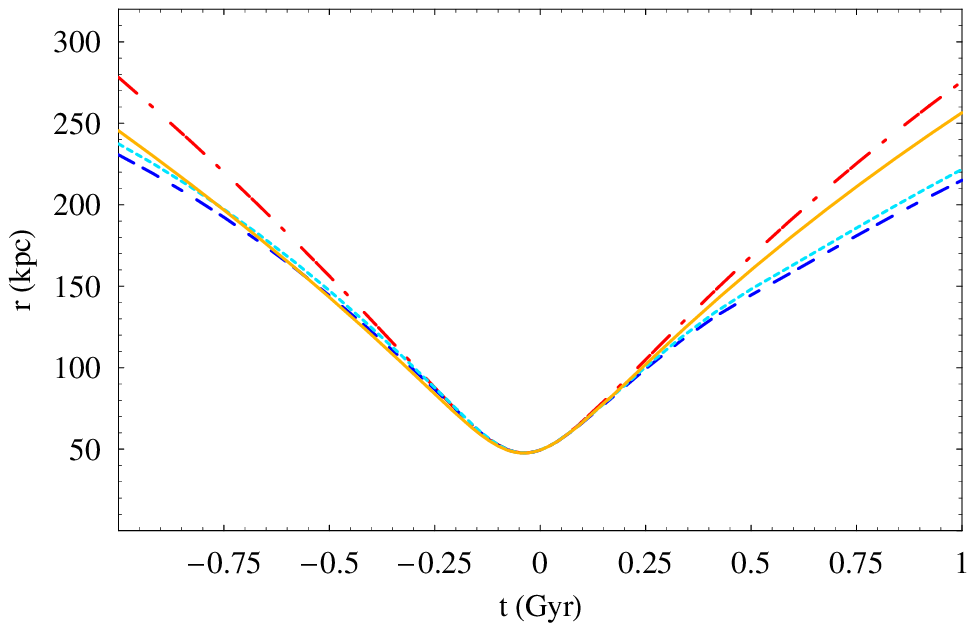}
\epsfysize= 4.75 cm\epsfbox{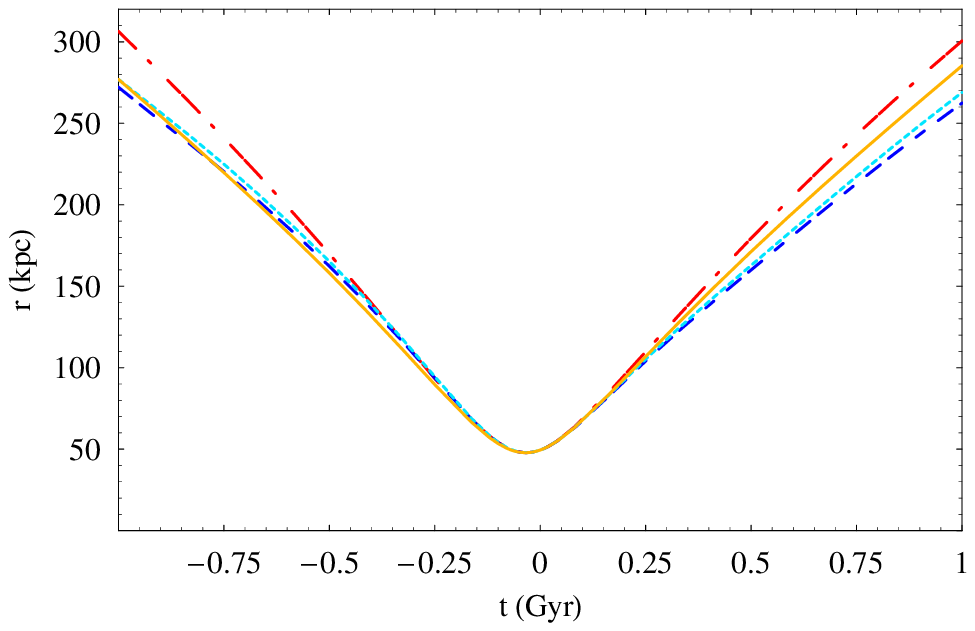}
      }
%
%
%
}
\caption{ {LMC: Galactocentric distance $r$, in kpc,  for $-1\leq t\leq 1$ Gyr. Red dash-dotted line: CDM. Blue dashed line: MOND ($\mu=X/(1+X)$). Light blue dotted line: MOND ($\mu=X$). Yellow continuous line: MOG. The initial condition for the position is $r=49.4$ kpc. Upper panel: for the velocity we adopted $\dot x_0 = -86+ 12=-74$ km s$^{-1}$, $\dot y_0 =-268+ 11=-257$ km s$^{-1}$, $\dot z_0 =252- 16=236$ km s$^{-1}$ yielding the minimum value $v=356.7$ km s$^{-1}$. Middle panel:  the central values of Table  \ref{LMCcoor}  have been adopted for the velocity. Lower panel: for the velocity we adopted $\dot x_0 = -86- 12=-98$ km s$^{-1}$, $\dot y_0 =-268- 11=-279$ km s$^{-1}$, $\dot z_0 =252+ 16=268$ km s$^{-1}$ yielding the maximum value $v=399.1$ km s$^{-1}$. For the masses of  MCs, entering their mutual interactions and the dynamical friction, both modelled in this integration, the total values (baryonic +  dark matter) have been adopted for CDM, while those encompassing only baryonic components have been used for MOG and MOND. }\label{dLMC}}
\end{figure}

\lb{grandicella}
In Figure~\ref{LMC_planes}  we show the sections in the coordinate planes of the LMC's orbits for CDM (red dash-dotted  curves), MOND (blue dashed lines and light blue dotted lines) and MOG (yellow continuous curves) over $-1\leq t\leq 1$ Gyr. The dynamical models and their parameters' values are those described in Section \ref{grandeee}.
%
%
%
%
We used \rfr{oleee}  for MOND ($A_{\rm ext}\ll A_{\rm N}$ and $A_{\rm ext}\ll A_0$) obtaining the blue dashed lines depicted; indeed, for LMC $A_{\rm N}/A_0=0.03$, so that \rfr{oleee} is  adequate for it by assuming $A_{\rm ext}=0.01 A_0$. Concerning the impact of EFE in MOND on LMC, we also investigated it in the case $A_{\rm ext}=A_{\rm N}$ and $A_{\rm ext}\ll A_0$; thus, we numerically integrated trajectories  with \rfr{stracaz} as well, which corresponds to an external field equal to the internal Newtonian one, obtaining the light blue dotted curves shown. The same approach will be used in Section \ref{piccina} for SMC.
%
%
%
%
The middle panel of Figure~\ref{dLMC} shows the Galactocentric distance of LMC for the central values of the velocity components of Table \ref{LMCcoor} over $-1\leq t\leq 1$ Gyr.
%
%
%
%

The smallest Galactocentric distance occurs for MOND, while MOG and CDM yield the largest one amounting to about $255-270$ kpc after $+1$ Gyr and $250-280$ kpc after $-1$ Gyr. MOG and CDM differ by about 15 kpc, while the discrepancy between MOG/CDM  and MOND is approximately $70-80$ kpc after $-1$ Gyr; in the past Gyr the discrepancy between MOG and CDM is of the order of 30 kpc, while MOND differs from MOG/CDM by a few 10 kpc. Over the next Gyr the Galactocentric distance of LMC  undergoes a steady increase. It maybe interesting to recall that, according to \citet{WuMag}, LMC is on a bound orbit; however, they did include neither the mutual interaction with SMC nor the dynamical friction.
The difference between the MOND trajectories for $\mu=X/(1+X)$ and $\mu=X$ is rather small;
discrepancies of the order of 10 kpc or less occur at $\pm1$ Gyr. All the models considered tend to undergo reciprocal departures after some about $\pm500$ Myr.
%
%
%
%

\begin{figure}
\centerline{\psfig{file=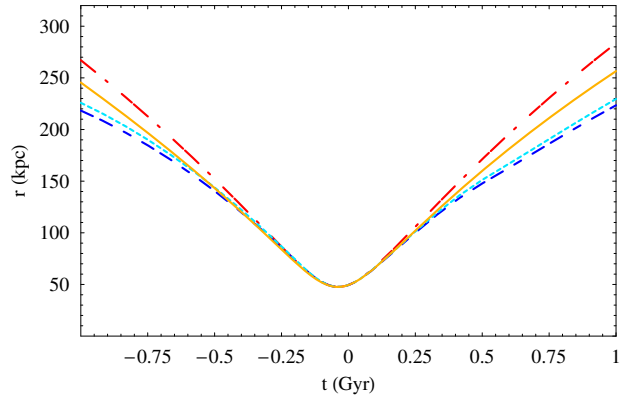,width=\columnwidth}}
\vspace*{8pt}
\caption{{LMC: Galactocentric distance $r$, in kpc,  for $-1\leq t\leq 1$ Gyr. Red dash-dotted line: CDM. Blue dashed line: MOND ($\mu=X/(1+X)$). Light blue dotted line: MOND ($\mu=X$). Yellow continuous line: MOG. The initial condition is $r=49.4$ kpc for the position;  for the velocity, the central values of Table ~\ref{LMCcoor} have been adopted. No dynamical friction has been applied in CDM and MOND. For the mass of SMC  the total value (baryonic +  dark matter) has been adopted for CDM, while that encompassing only baryonic components has been used for MOG and MOND. The time span of the integration is $-1\leq t\leq 1$ Gyr.}\label{dLMC_nodynfric}}
\end{figure}

Figure \ref{dLMC_nodynfric}
shows the impact of the dynamical friction; after $+1$ Gyr, without modeling it in CDM and MOND, the mutual difference between CDM and MOG tends to increase by about 10 kpc, while the MONDian trajectories are left almost unaffected. Instead, at $-1$ Gyr the discrepancy between CDM and MOG gets reduced by 10 kpc, while the MOND distance is smaller by about 20 kpc.

The impact of the uncertainties in the velocity components of LMC has been evaluated as it will be done for SMC in Section \ref{piccina}; it is shown in the upper (minimum velocity) and lower (maximum velocity) panels of Figure \ref{dLMC}. Differences with respect to the nominal case are present. Indeed, for the smallest value of the velocity ($\dot x_0 = -86+ 12=-74$ km s$^{-1}$, $\dot y_0 =-268+ 11=-257$ km s$^{-1}$, $\dot z_0 =252- 16=236$ km s$^{-1}$ ), the overall discrepancy among CDM/MOG and MOND is of the order of $50-60$ kpc after $+1$ Gyr, with a  reduction of the final distances in CDM/MOG with respect to the middle panel of Figure \ref{dLMC} (20 kpc for CDM, 30 kpc for MOG and 50 kpc for MOND); after $-1$ Gyr the CDM distance is 30 kpc smaller than for the nominal values of the velocity components, MOG is 35 kpc below the level of the middle panel of Figure \ref{dLMC}, while the MOND curves experience a reduction of about 40 kpc. For $\dot x_0 = -86- 12=-98$ km s$^{-1}$, $\dot y_0 =-268- 11=-279$ km s$^{-1}$, $\dot z_0 =252+ 16=268$ km s$^{-1}$, corresponding to the maximum velocity, the relative discrepancy after $+1$ Gyr among the various models is about $60-80$ kpc, with an increase of each of them with respect to the middle panel of Figure \ref{dLMC} (25 kpc for CDM, 30 kpc for MOG, and $50-60$ kpc for MOND); also at $-1$ Gyr there is an overall increase with respect to the case of the nominal values of the velocity components (20 kpc for CDM, 30 kpc for MOG, and $30-40$ kpc for MOND).
\subsection{The Small Magellanic Cloud}
\begin{figure}
\centerline{
\vbox{
\epsfysize= 5.75 cm\epsfbox{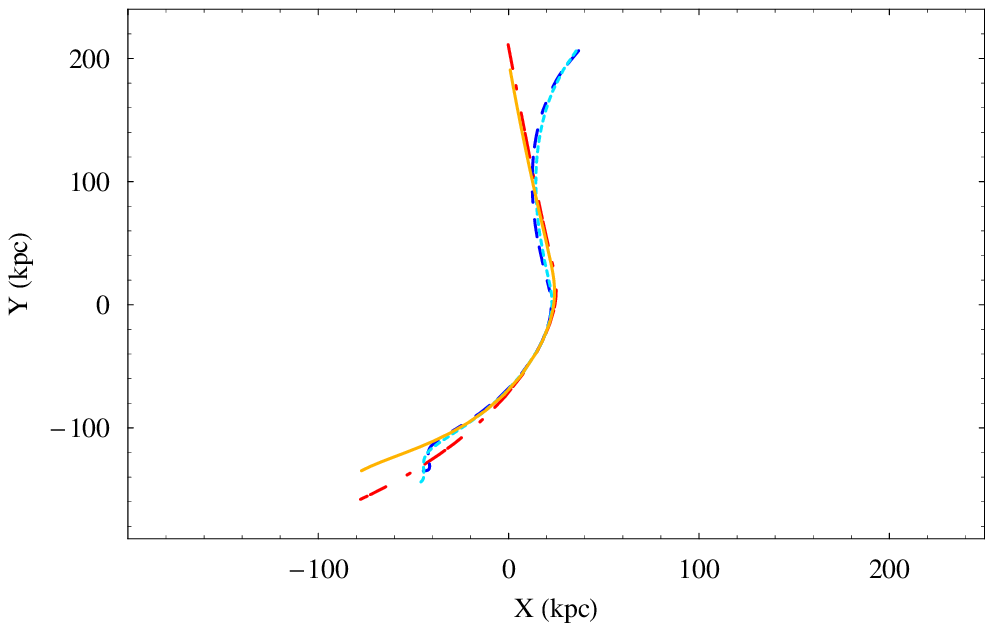}
\epsfysize= 5.75 cm\epsfbox{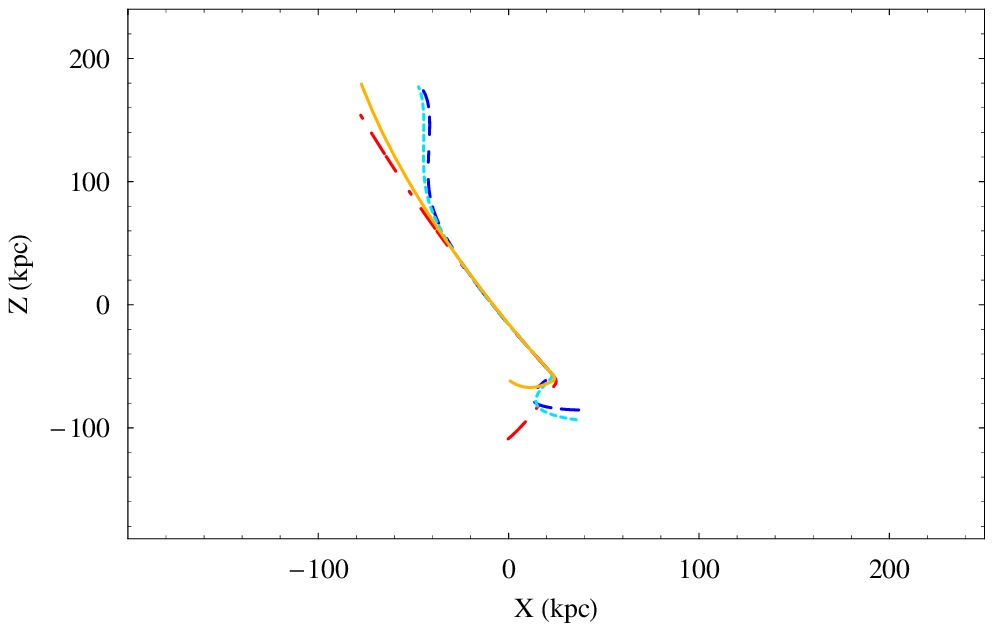}
\epsfysize= 5.75 cm\epsfbox{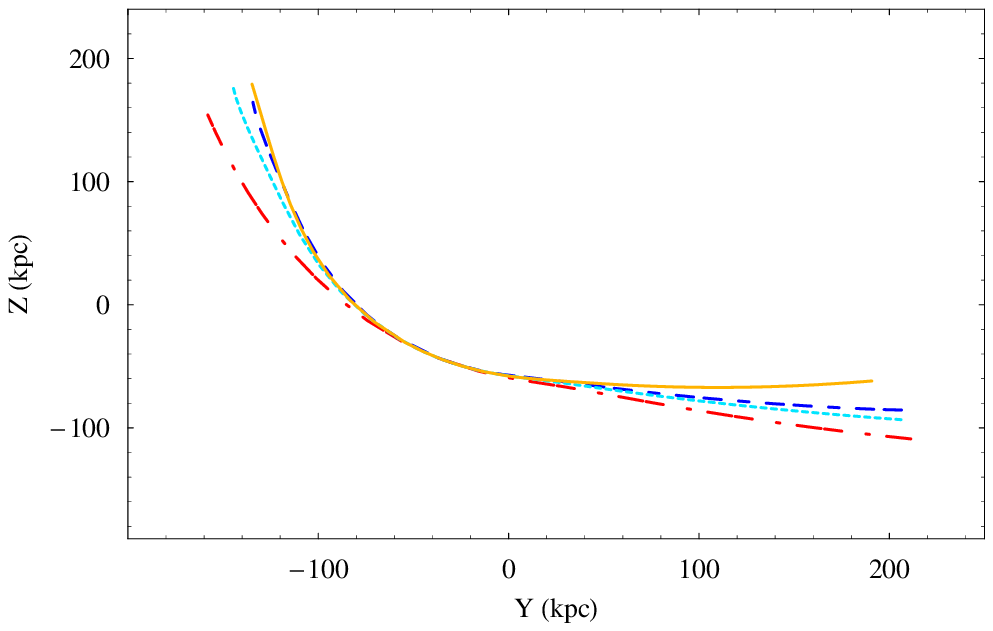}
      }
%
%
%
}
\caption{ {Sections in the coordinate planes of the numerically integrated trajectories of  SMC experiencing: a) The Newtonian acceleration with CDM (red dash-dotted  line) b)  The MOND acceleration with $\mu=X/(1+X)$   (blue dashed  line) c) The MOND acceleration with $\mu=X$   (light blue dotted line) d) The MOG acceleration (yellow continuous  line). The central values of the initial conditions of Table  \ref{SMCcoor} have been used. For the baryonic masses of  MW's bulge and  disk  we used the values by \protect{\citet{McG}}, with a total baryonic mass of $M=6.5\times 10^{10}$ M$_{\odot}$.  For the masses of MCs, entering their mutual interactions and the dynamical friction, both modelled in this integration, the total values (baryonic +  dark matter) have been adopted for CDM, while those encompassing only baryonic components have been used for MOG and MOND. have been adopted. The time span of the integration is $-1\leq t\leq 1$ Gyr.}    \label{SMC_planes}
}
\end{figure}

\lb{piccina}
Figure~\ref{SMC_planes}  depicts the sections in the coordinate planes of the SMC's orbits over $-1\leq t\leq 1$ Gyr for CDM (red dash-dotted  curves), MOND (blue dashed lines and light blue dotted lines) and MOG (yellow continuous curves) by using the central values of the initial velocities of Table   \ref{SMCcoor} and the same values of Section \ref{grandicella} for the masses of MW and  MCs and of the other models' parameters.
Concerning MOND and the impact of EFE, we followed the same approach as for LMC in Section \ref{grandicella}.
It can be noted that  MOND, MOG and CDM yield  different orbital patterns, especially in the $\{xy\}$ and $\{xz\}$ planes, and it is possible, in principle, to discriminate among them.
%
%
%
%
%
%
%
%
%
\begin{figure}
\centerline{
\vbox{
\epsfysize= 4.75 cm\epsfbox{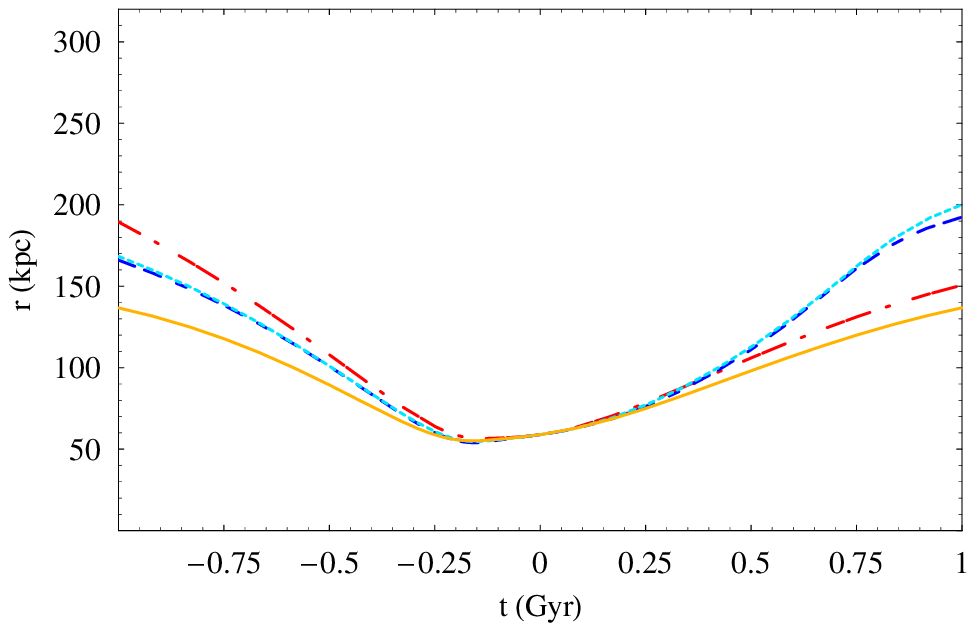}
\epsfysize= 4.75 cm\epsfbox{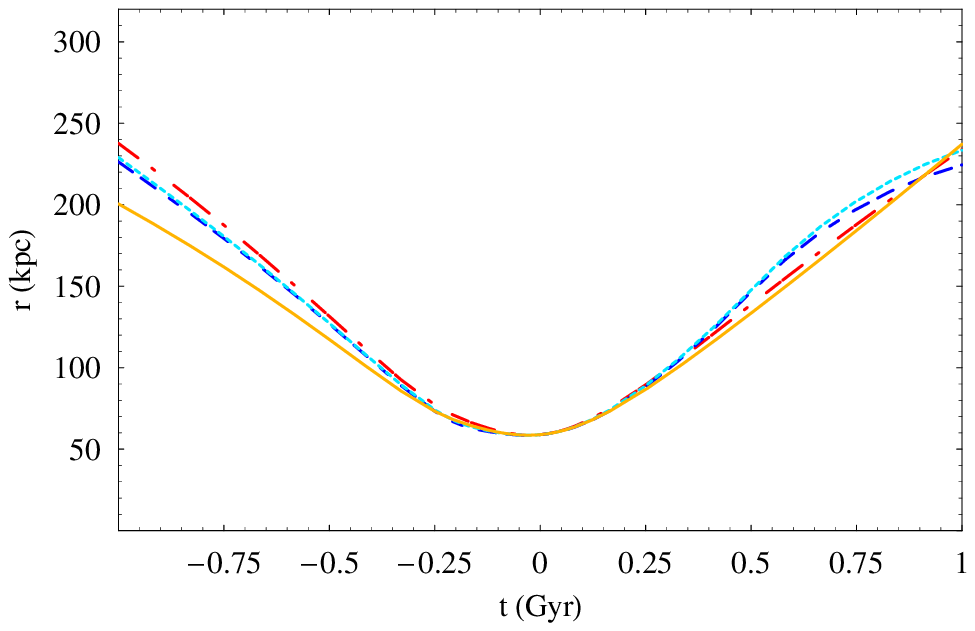}
\epsfysize= 4.75 cm\epsfbox{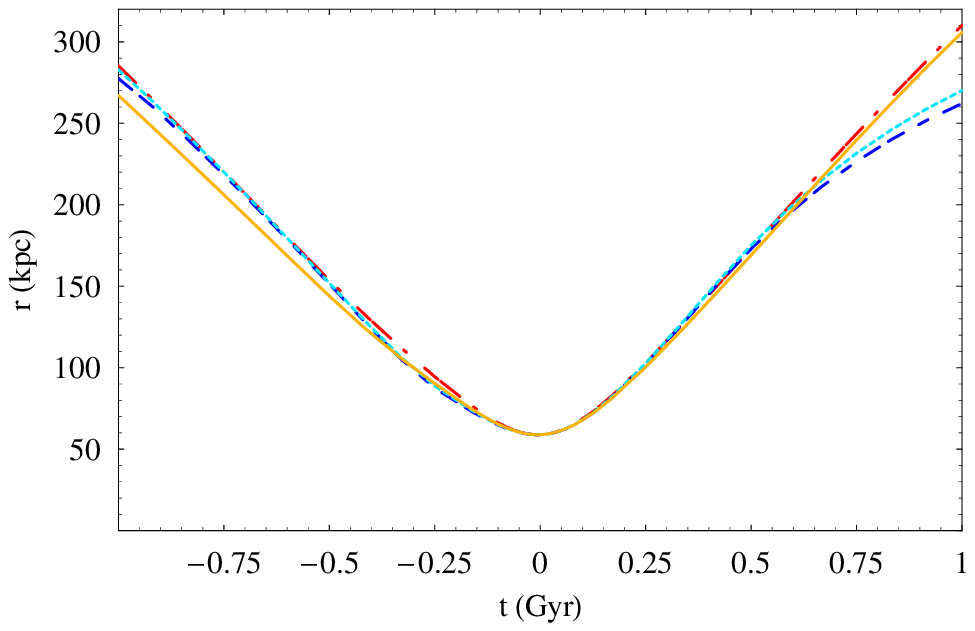}
      }
%
%
%
}
\caption{ {SMC: Galactocentric distance $r$, in kpc,  for $-1\leq t\leq 1$ Gyr. Red dash-dotted line: CDM. Blue dashed line: MOND ($\mu=X/(1+X)$). Light blue dotted line: MOND ($\mu=X$). Yellow continuous line: MOG. The initial condition for the position is $r=58.9$ kpc. Upper panel: for the velocity we adopted $\dot x_0 = -87+ 48=-39$ km s$^{-1}$, $\dot y_0 =-247+ 42=-205$ km s$^{-1}$, $\dot z_0 =149- 37=112$ km s$^{-1}$ yielding the minimum value $v=236$ km s$^{-1}$. Middle panel:  the central values of Table  \ref{SMCcoor}  have been adopted for the velocity. Lower panel: for the velocity we adopted $\dot x_0 = -87- 48=-135$ km s$^{-1}$, $\dot y_0 =-247- 42=-289$ km s$^{-1}$, $\dot z_0 =149+ 37=186$ km s$^{-1}$ yielding the maximum value $v=369$ km s$^{-1}$. For the masses of  MCs, entering their mutual interactions and the dynamical friction, both modelled in this integration, the total values (baryonic +  dark matter) have been adopted for CDM, while those encompassing only baryonic components have been used for MOG and MOND. }\label{altra}}
\end{figure}

In the middle panel of Figure~\ref{altra}  we plot the time evolution of the Galactocentric distance  of SMC according to CDM, MOND and MOG in the next Gyr for the central values of the velocity components of Table \ref{SMCcoor}.
The distance reached in all the three models after $+1$ Gyr is practically the same, amounting to $220-230$ kpc.
After $-1$ Gyr the scatter among the models considered is larger, amounting to about 40 kpc.
As for LMC, all the models considered tend to undergo reciprocal departures after some about $\pm500$ Myr, although smaller.
%

\begin{figure}
\centerline{\psfig{file=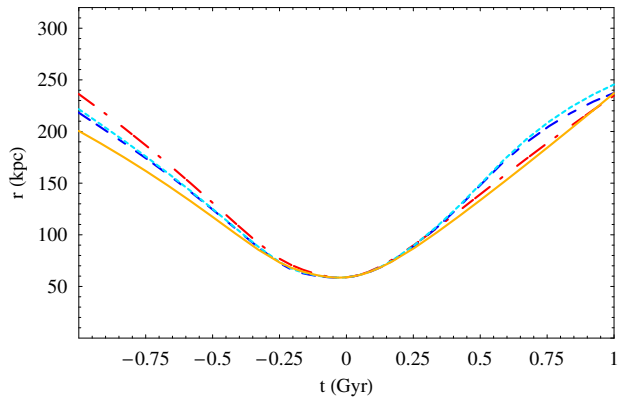,width=\columnwidth}}
\vspace*{8pt}
\caption{{SMC: Galactocentric distance $r$, in kpc,  for $-1\leq t\leq 1$ Gyr. Red dash-dotted line: CDM. Blue dashed line: MOND ($\mu=X/(1+X)$). Light blue dotted line: MOND ($\mu=X$). Yellow continuous line: MOG. The initial condition is $r=49.4$ kpc for the position;  for the velocity, the central values of Table ~\ref{SMCcoor} have been adopted. No dynamical friction has been applied in CDM and MOND. For the mass of LMC  the total value (baryonic +  dark matter) has been adopted for CDM, while that encompassing only baryonic components has been used for MOG and MOND. }\label{dSMC_nodynfric}}
\end{figure}

Figure \ref{dSMC_nodynfric} shows that switching off the dynamical friction in CDM and MOND does not substantially alter the overall picture.
%
%

The uncertainty in the velocity components of SMC may have different consequences on its orbit for the models considered. The Galactocentric distance of SMC  for the maximum value of its speed, i.e. $v_{\rm SMC}=369$ km s$^{-1}$ corresponding to
$\dot x_0 = -87- 48=-135$ km s$^{-1}$, $\dot y_0 =-247- 42=-289$ km s$^{-1}$, $\dot z_0 =149+ 37=186$ km s$^{-1}$,
is shown in the lower panel of Figure~\ref{altra}. By comparing it with the middle panel of Figure~\ref{altra}, it can be noted that, after $+1$ Gyr, the Galactocentric distance increases, in particular in CDM and MOG; instead, at $-1$ Gyr the increase is more uniform for all the models.
In the upper panel of Figure~\ref{altra}  we depict the case for  $\dot x_0 = -87+ 48=-39$ km s$^{-1}$, $\dot y_0 =-247+ 42=-205$ km s$^{-1}$, $\dot z_0 =149- 37=112$ km s$^{-1}$ yielding the minimum value for the SMC's speed $v=236$ km s$^{-1}$.
In this case, CDM and MOG yield a smaller Galactocentric distance after $+1$ Gyr: indeed,  it is as large as $140-150$ kpc. Instead, after  $-1$ Gyr the MOND curves lie in between the MOG/CDM ones, with an overall reduction of $50-60$ kpc for all the models. 
%
%
\subsection{The  mutual distance between SMC and LMC}
\begin{figure}
\centerline{
\vbox{
\epsfysize= 5.5 cm\epsfbox{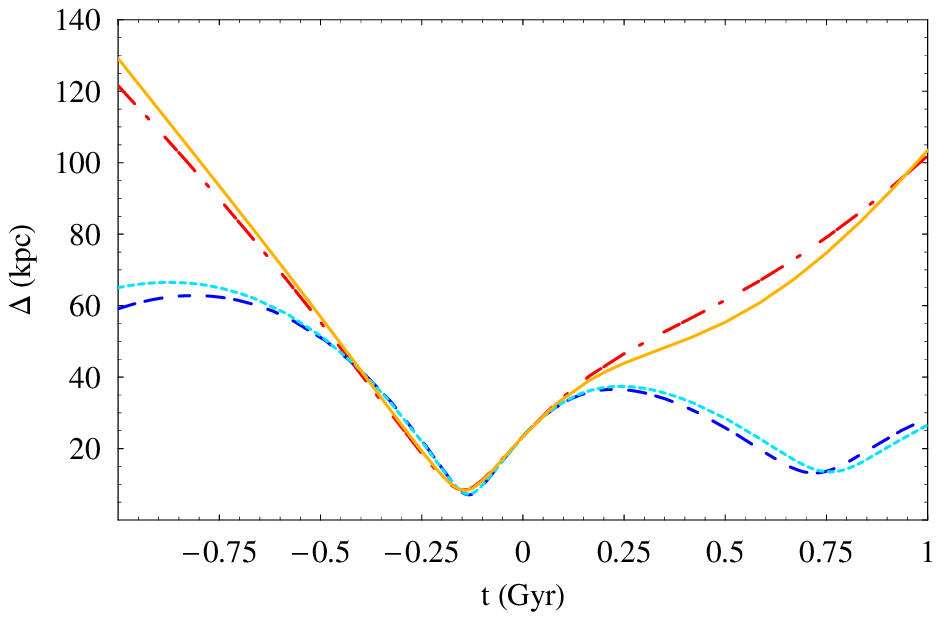}
\epsfysize= 5.5 cm\epsfbox{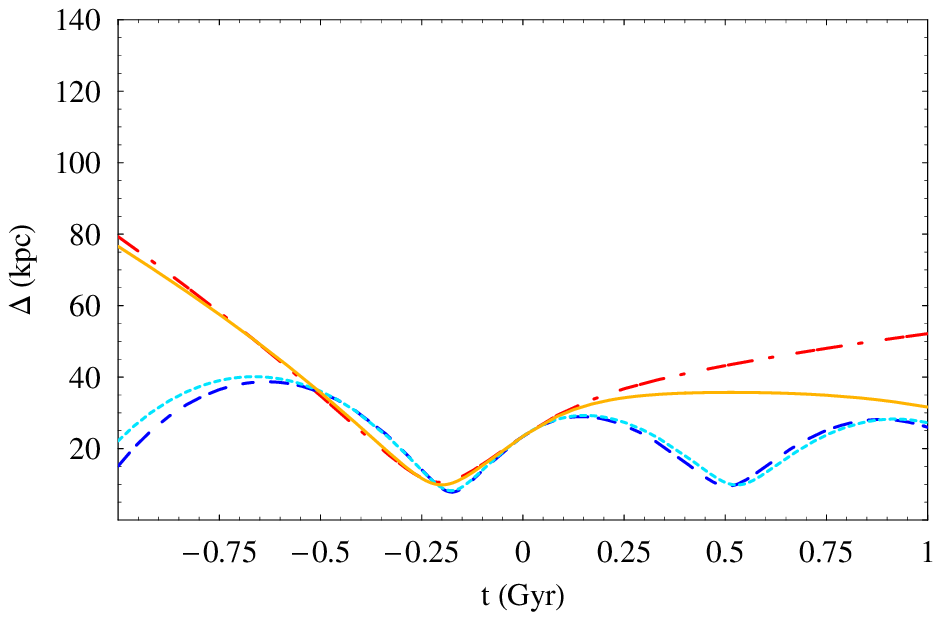}
\epsfysize= 5.5 cm\epsfbox{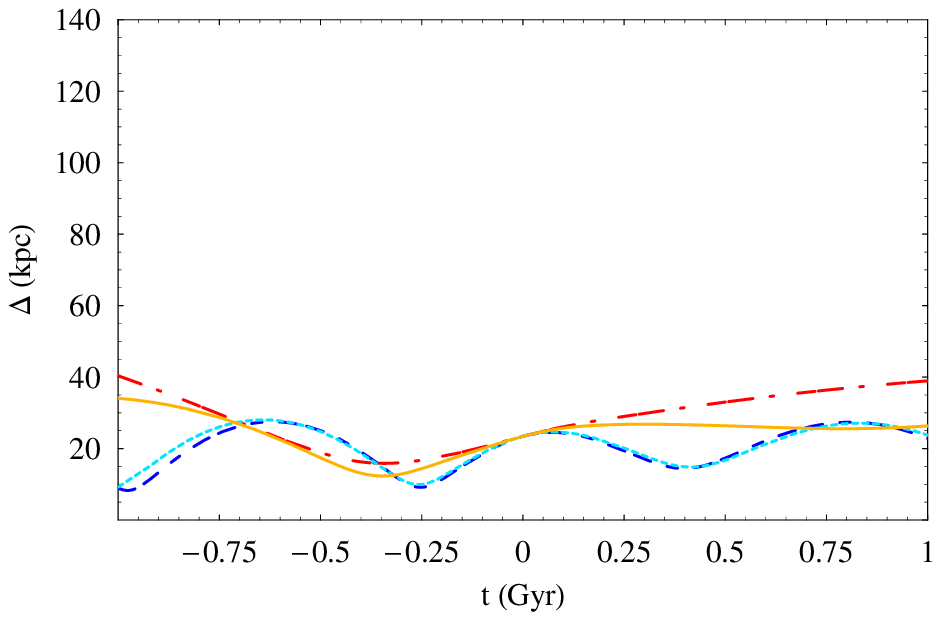}
      }
%
%
%
}
\caption{{Mutual distance $\Delta$ between SMC and LMC, in kpc,  for $-1\leq t\leq 1$ Gyr. Red dash-dotted line: CDM. Blue dashed line: MOND ($\mu=X/(1+X)$). Light blue dotted line: MOND ($\mu=X$). Yellow continuous line: MOG. Upper panel: for the velocity components of both SMC and LMC we used their minimum values. Middle panel: for the velocity components of both SMC and LMC we used their central values. Lower panel: for the velocity components of both SMC and LMC we used their maximum values. For the masses of  MCs, entering their mutual interactions and the dynamical friction, the total values (baryonic +  dark matter) have been adopted for CDM, while those encompassing only baryonic components have been used for MOG and MOND.}\label{mutual}}
\end{figure}

In the middle panel of Figure~\ref{mutual} the mutual SMC$-$LMC distance $\Delta=\sqrt{(x_{\rm LMC}-x_{\rm SMC})^2+(y_{\rm LMC}-y_{\rm SMC})^2+(z_{\rm LMC}-z_{\rm SMC})^2}$ is shown for the central values of the velocity components of both MCs an with the same models and parameters' values of Section \ref{grandeee}.
The pattern by CDM is quite different with respect to those by MOND and, to a lesser extent, MOG, both in the size of the distance reached  and, especially, in the temporal signature. After $+1$ Gyr, MOND exhibits a bounce yielding the smallest maximum reciprocal separation, i.e. about 25 kpc, while for CDM, which yields an increasing signal, it is approximately 50 kpc. Instead, at $-1$ Gyr CDM and MOG reach 80 kpc, while MOND is around 20 kpc.

\begin{figure}
\centerline{\psfig{file=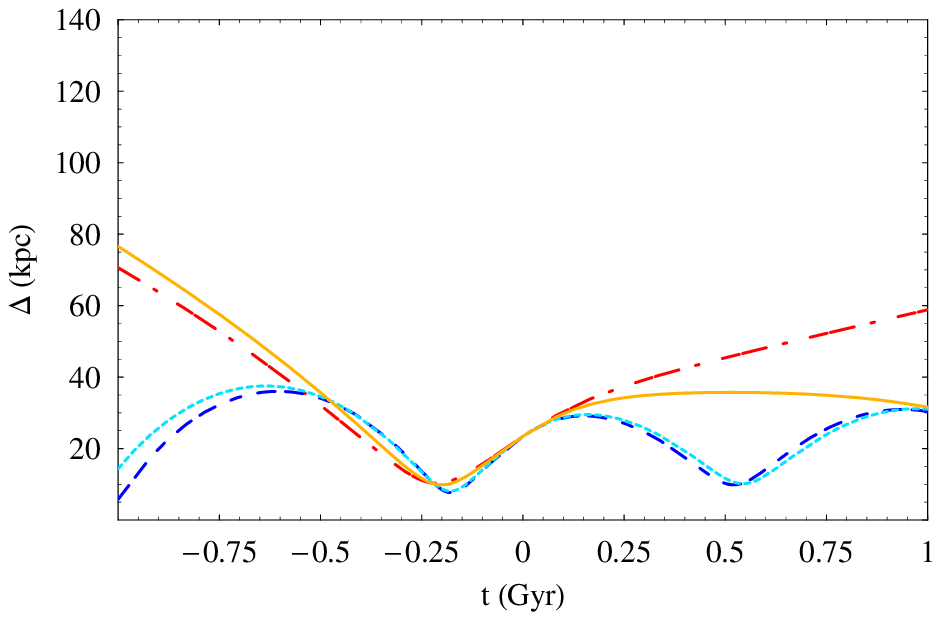,width=\columnwidth}}
\vspace*{8pt}
\caption{{Mutual distance $\Delta$ between SMC and LMC, in kpc,  for $-1\leq t\leq 1$ Gyr. Red dash-dotted line: CDM. Blue dashed line: MOND ($\mu=X/(1+X)$). Light blue dotted line: MOND ($\mu=X$). Yellow continuous line: MOG. For the velocity components of both SMC and LMC we used their central values. No dynamical friction has been modeled. For the masses of  MCs, entering their mutual interaction, the total values (baryonic +  dark matter) have been adopted for CDM, while those encompassing only baryonic components have been used for MOG and MOND.}\label{dMC_nodynfric}}
\end{figure}

In Figure \ref{dMC_nodynfric} we show  MCs mutual distance without dynamical friction. After $+1$ Gyr, the CDM maximum distance is 60 kpc, while the MOND curves tend to approach the MOG one at 30 kpc. After $-1$ Gyr, CDM reaches about 70 kpc; MOND is below the 20 kpc level.

The impact of the uncertainties in the velocity of both SMC and LMC on the mutual separation is depicted in the upper and lower panels of Figure~\ref{mutual}.
Low velocities (upper panel) yield a change of the bouncing time for MOND  and an increase of $\Delta$ in CDM and MOG by more than 50 kpc for $\pm 1$ Gyr. Note also the increase of the MOND curves at $-1$ Gyr: they pass from 20 kpc to 60 kpc, while they are not substantially changed after $+1$ Gyr. On the contrary, high velocities (lower panel) tend to yield an overall reduction of the distances among all the models, with $\Delta_{\rm CDM}$ reduced down to 40 kpc.

Generally speaking, the repeated close encounters (in MOND and MOG) may have an impact on the star formation history and/or morphology of both MCs; anyway, discussing such  interesting issues is beyond the scope of the present paper.
\section{Summary and conclusions}
We simultaneously  integrated in a numerical way the orbits of both MCs for $-1\leq t\leq 1$ Gyr within MOND, MOG and CDM  to see if, at least in principle, it is possible to discriminate among them. This is, in principle, important also because it is believed that MS  follows the orbits of MCs.

 Since LMC and SMC are  at about 50-60 kpc from the Galactic center, they are ideal candidates to investigate the deep MOND regime ($A_{\rm N}/A_0=0.03-0.02$); moreover, the details of the realistic mass distribution can be neglected. Thus, for MOND (and MOG) we used a pointlike approximation for the baryonic mass of MW; we tested it by successfully reproducing the orbital paths of the Sagittarius dwarf galaxy ($r=17$ kpc) obtained by other researchers with the MONDian fully non-linear modified Poisson equation. For CDM we used a logarithmic halo potential which is able to reproduce the value of the Galactic mass at 60 kpc obtained independently by analyzing different tracers. We also took into account the mutual MCs interaction and, for MOND and CDM, the dynamical friction as well. For the masses of MCs we used the total (baryonic $+$ dark matter) values dynamically inferred
 in CDM, and the smaller ones coming from direct detection of the electromagnetic radiation emitted by  stars and neutral gas in MOND and MOG.

It turns out that, in fact, CDM, MOND and MOG do yield different trajectories for SMC and LMC. In general, the spatial extension of the orbits' sections in the coordinate planes is larger for CDM and MOG with respect to MOND. SMC experiences larger discrepancies among the various models than LMC.
Since for  MW  $A_{\rm ext}\approx 0.01 A_0$, we also investigated  EFE in MOND by considering not only  $A_{\rm ext}\ll A_{\rm N}, A_{\rm ext}\ll A_0$, but also  $A_{\rm ext}\approx A_{\rm N}, A_{\rm ext}\ll A_0$ which cannot be excluded in view of the lingering uncertainty in  MW's EFE. The resulting orbital patterns are rather similar to those obtained neglecting $A_{\rm ext}$.
We also investigated the impact of the present-day uncertainties in the velocity components of MCs on their trajectories in the models considered by finding that it is more notable for SMC than LMC;  CDM and MOG are more sensitive to such a source of bias than MOND. In general, the largest discrepancies among the various models occur around $\pm1$ Gyr. This suggests that extending the integration time may yield interesting findings; it may be the subject of further analyses. Over the timescale considered, the dynamical friction does not make the paths too much different in the various models examined.

\section*{Acknowledgments}
I gratefully thank an anonymous referee for her/his remarkable continuous efforts to improve the manuscript with important remarks.


\end{document}